\documentclass[%
preprint,
superscriptaddress,
amsmath,amssymb,
aps,
]{revtex4-1}

\usepackage{graphicx}
\usepackage{dcolumn}
\usepackage{bm}

\usepackage[T1]{fontenc}       
\usepackage[utf8]{inputenc}
\usepackage{lmodern}%
\usepackage{hyperref}%
\usepackage{xcolor}%
\usepackage{placeins}
\usepackage{tabularx}
\usepackage{booktabs}

\usepackage[normalem]{ulem}

\newcommand{\beq}{\begin{equation}\begin{aligned}}
\newcommand{\eeq}{\end{aligned}\end{equation}}

\newcommand{\mum}{$\mu$m}
\newcommand{\ws}{WS\ensuremath{_2}}
\newcommand{\wse}{WSe\ensuremath{_2}}
\newcommand{\mos}{MoS\ensuremath{_2}}
\newcommand{\mose}{MoSe\ensuremath{_2}}

\newcommand{\degree}{\ensuremath{^\circ\,}}

\newcommand{\gam}{\ensuremath{\Gamma}}

\newcommand{\eg}{\emph{e.g.}}

\newcommand{\dqmp}{Department of Quantum Matter Physics, University of Geneva, 24 Quai Ernest Ansermet, CH-1211 Geneva, Switzerland}
\newcommand{\gap}{Group of Applied Physics, University of Geneva, 24 Quai Ernest Ansermet, CH-1211 Geneva, Switzerland}
\newcommand{\mancha}{National Graphene Institute, University of Manchester, Booth St E, M13 9PL, Manchester, UK}
\newcommand{\manchb}{School of Physics \& Astronomy, University of Manchester, Oxford Road, M13 9PL, Manchester, UK}
\newcommand{\manchc}{Henry Royce Institute for Advanced Materials, M13 9PL, Manchester, UK}
\newcommand{\hbngrow}{National Institute for Materials Science, 1-1 Namiki, Tsukuba, 305-0044, Japan}
\newcommand{\nott}{School of Physics \& Astronomy, The University of Nottingham, Nottingham, NG7 2RD, UK}
\newcommand{\ukraine}{Institute for Problems of Materials Science, NAS of Ukraine, Chernivtsi Branch, 5 I. Vilde Str., Chernivtsi 58001, Ukraine}

\definecolor{linkcol}{rgb}{0,0,0.4}
\definecolor{citecol}{rgb}{0.5,0,0}

\hypersetup{
	bookmarksopen=true,
	pdftitle="Design of van der Waals Interfaces for Broad-Spectrum Optoelectronics",
	pdfauthor="Ubrig et al",
	pdfsubject="Design of van der Waals Interfaces for Broad-Spectrum Optoelectronics", 
	pdfstartview={FitH},    
	pdfmenubar=true, 
	pdfhighlight=/O, 
	colorlinks=true, 
	pdfpagemode=UseNone, 
	pdfpagelayout=SinglePage, 
	pdffitwindow=true, 
	linkcolor=linkcol, 
	citecolor=linkcol, 
	urlcolor=blue
}

\begin{document}
	
	
	\title{Design of \emph{van der Waals} Interfaces for Broad-Spectrum Optoelectronics}
	
	%
	%
	
	\author{Nicolas Ubrig}
	\email{nicolas.ubrig@unige.ch}
	\author{Evgeniy Ponomarev}
	\affiliation{\dqmp}
	\affiliation{\gap}
	\author{Johanna Zultak} 
	\affiliation{\mancha}
	\affiliation{\manchb}
	\affiliation{\manchc}
	\author{Daniil Domaretskiy}
	\affiliation{\dqmp}
	\affiliation{\gap}
	\author{Viktor Zólyomi}	
	\affiliation{\mancha}
	\author{Daniel Terry}		
	\author{James Howarth}	
	\affiliation{\mancha}
	\affiliation{\manchb}
	\affiliation{\manchc}
	\author{Ignacio Gutiérrez-Lezama}
	\affiliation{\dqmp}
	\affiliation{\gap}
	\author{Alexander Zhukov}	
	\affiliation{\mancha}
	\affiliation{\manchb}
	\affiliation{\manchc}
	\author{Zakhar R. Kudrynskyi}	
	\affiliation{\nott}
	\author{Zakhar D. Kovalyuk}	
	\affiliation{\ukraine}
	\author{Amalia Patanè}
	\affiliation{\nott}	
	\author{Takashi Taniguchi}
	\author{Kenji Watanabe}
	\affiliation{\hbngrow}
	\author{Roman V. Gorbachev}	
	\author{Vladimir I. Fal'ko}		
	\email{Vladimir.Falko@manchester.ac.uk}
	\affiliation{\mancha}
	\affiliation{\manchb}
	\affiliation{\manchc}
	\author{Alberto F. Morpurgo}
	\email{alberto.morpurgo@unige.ch}
	\affiliation{\dqmp}
	\affiliation{\gap}
	%
	
	
	\date{\today}
	
	
	\maketitle

	{\bfseries Van der Waals (vdW) interfaces based on two dimensional (2D) materials are promising for optoelectronics, as interlayer transitions between different compounds allow tailoring the spectral response over a broad range. However, issues such as lattice mismatch or a small misalignment of the constituent layers can drastically suppress electron-photon coupling for these interlayer transitions. Here, we engineer type-II interfaces by assembling atomically thin crystals that have the bottom of the conduction band and the top of the valence band at the \gam-point, thus avoiding any momentum mismatch. We find that these vdW interfaces exhibit radiative optical transitions irrespective of lattice constant, rotational/translational alignment of the two layers, or whether the constituent materials are direct or indirect gap semiconductors. Being robust and of general validity, our results broaden the scope of future optoelectronics device applications based on two-dimensional materials.\\}

	\emph{Van der Waals} interfaces of interest for optoelectronics consist of two distinct layered semiconductors with a suitable energetic alignment of their conduction and valence bands, such that electron and hole excitations reside in the two separate layers.\cite{geim_van_2013,novoselov_2d_2016,fang_strong_2014,rivera_interlayer_2018} This allows the interfacial band gap to be controlled by material selection --as well as by application of an electrical bias or strain\cite{castellanos-gomez_local_2013,rivera_observation_2015,ross_interlayer_2017,mennel_second_2018,ciarrocchi_polarization_2019}-- so that electron-hole recombination across the layers generates photons with frequency determined over a broad range at the design stage. Choosing the interface components among the vast gamut of 2D materials --including semiconducting transition metal dichalcogenides (TMDs, MoS$_2$, \mose{}, MoTe$_2$, \ws{}, WSe$_2$, ReS$_2$, ZrS$_2$, etc.), III-VI compounds (InSe, GaSe), black phosphorous, and even magnetic semiconductors (CrI$_3$, CrCl$_3$, CrBr$_3$, etc.)-- enables, at least in principle, to cover a spectral range from the far infra-red to the violet. In practice, however, efficient light-emission from interlayer recombination requires the corresponding electron-hole transition to be direct in reciprocal (k-) space: the bottom of the conduction band in one layer has to be centered in k-space at the same position as the top of the valence band in the other layer.\cite{yu_anomalous_2015} This requirement poses severe constraints as concluded from heterostructures of monolayer semiconducting TMDs, the systems that have been so far mostly used to realize light-emitting vdW interfaces.\cite{lee_atomically_2014,furchi_photovoltaic_2014,mueller_2d_2015,ross_interlayer_2017,binder_upconverted_2019} Indeed, in this case the minimum of the conduction band and top of valence band are at the K/K’ points in the Brillouin zone and the presence of radiative interlayer transitions requires combining compounds with both matched lattice constants and virtually perfect rotational alignment.\cite{yu_anomalous_2015,zhu_interfacial_2017} If not, the k-space mismatch between the K/K’ points in the two materials prevents interlayer radiative recombination.\cite{ponomarev_semiconducting_2018,binder_upconverted_2019} To bypass these limitations, it is important to identify a mechanism enabling the occurrence of robust radiative transitions in vdW interfaces, as well as classes of semiconducting 2D materials to implement it.\\
	
	Here we propose a general strategy to form vdW interfaces of 2D semiconductors supporting interlayer transitions that are direct in k-space, irrespective of the crystal symmetry, lattice constant, or crystallographic alignment of the constituent materials. If the materials forming the interface are selected so that the conduction band minimum in one and the valence band maximum in the other are centered at the \gam{}-point of reciprocal space, interlayer transitions will be direct in k-space as long as the energetic alignment of the bands is of type II (because the \gam{}-point resides at k = 0, and hence coincides for all materials). To demonstrate this strategy, we use bilayers and thicker multilayers of different TMDs (\ws{}, MoS$_2$, and MoSe$_2$) having the maximum of their valence band at the \gam{}-point,\cite{splendiani_emerging_2010} and show that they enable direct interlayer transition in vdW interfaces with InSe multilayers, which have their conduction band minimum at \gam{}.\cite{mudd_tuning_2013,bandurin_high_2017} Light emission from \gam{}-point interlayer transitions is well known in covalently bonded heterostructures of GaAs/AlGaAs  and CdTe/ZnSe,\cite{o._gobel_fabrication_1990,he_mbe_1990,odonnell_zncdsse_1992,yu_band_1992,butov_photoluminescence_1999} which form the basis for a multitude of technological applications including light-emitting diodes, different types of lasers, radiation detectors, etc.\cite{kroemer_nobel_2001,klingshirn_semiconductor_2012,butov_excitonic_2017} The advantage of vdW interfaces is that the constituent materials neither need to be lattice matched nor satisfy any other stringent conditions, broadening the choice of materials that can be used and, correspondingly, the range of accessible photon frequencies.\\
	
\begin{figure*}
		\centering
		\includegraphics[width=.51\linewidth]{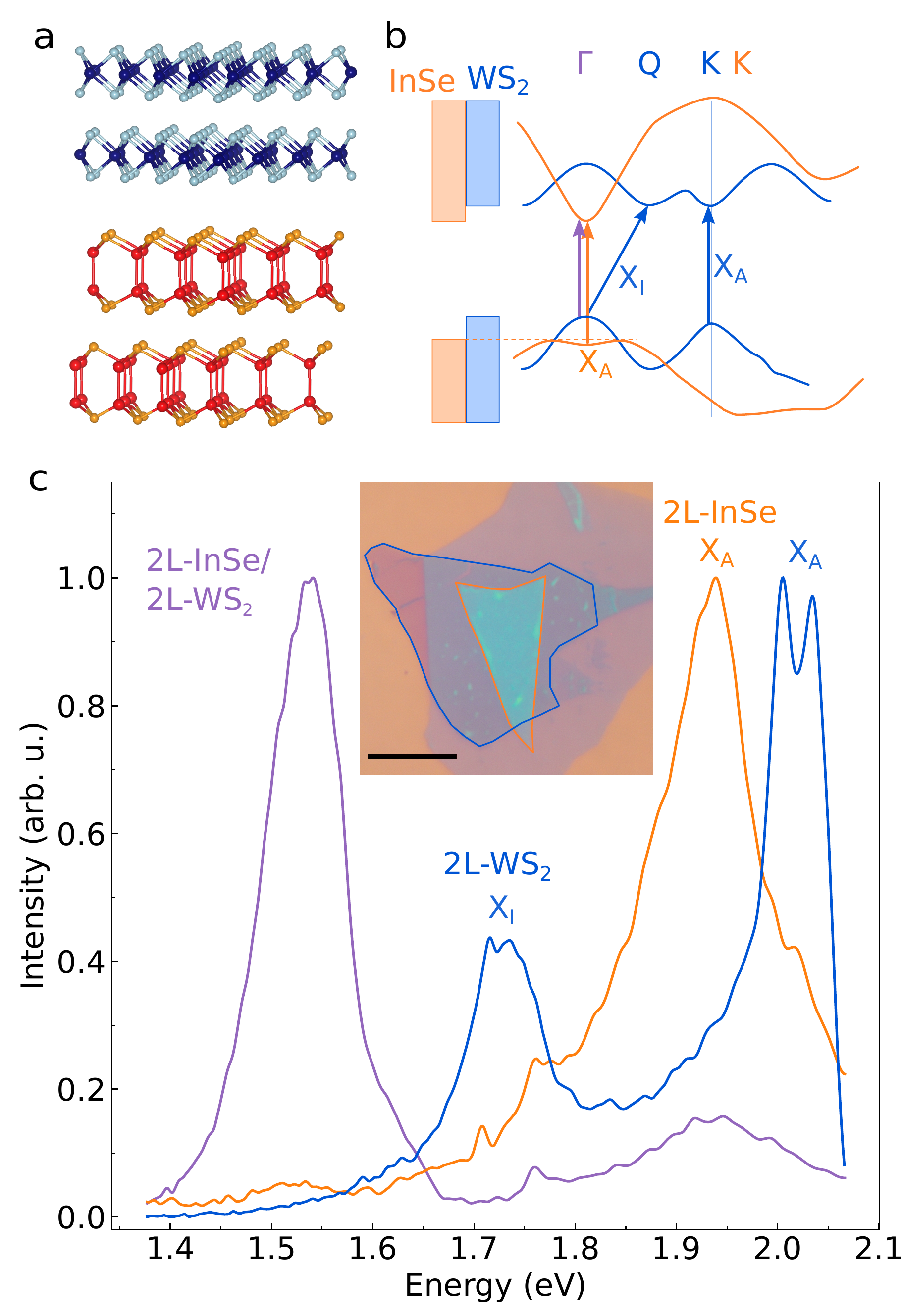}
		\caption{\textbf{Photoluminescence of 2L-InSe/2L-\ws{} interfaces.} \textbf{\sffamily a}. Schematics of a 2L-InSe/2L-\ws{} interface with the \ws\ bilayer on top and the InSe on the bottom. These two systems have a honeycomb lattice structure with lattice constants differing by approximatively 15\%. The relevant parts of the band structure of the two layers are shown in \textbf{\sffamily b} (the orange and blue lines represent the conduction and valence band edge of 2L-InSe and of 2L-\ws{}, respectively). The blue and the orange arrows indicate the transitions observed in the PL of the individual bilayers; the purple arrow represents the interlayer transition from the conduction band of 2L-InSe to the valence band of 2L-\ws\ that is observed in the interfacial PL. \textbf{\sffamily c}. Measured photoluminescence spectra of bare 2L-InSe (orange line), bare 2L-\ws\ (blue line), and of their interface (purple line). The labels on the PL curves ({\sffamily X$_{\mathsf A}$} and {\sffamily X$_{\mathsf I}$}) measured on the individual InSe and \ws\ bilayers refer to the transitions pointed to by the arrows in panel \textbf{\sffamily b}. Finding that the interfacial PL peaks at an energy smaller than that of all transitions occurring in the individual constituent layers indicates that the interfacial PL originates from an interlayer transition. The inset shows an optical microscope image of a h-BN-encapsulated (hexagonal-Boron Nitride) 2L-InSe/2L-\ws{} interface (the scale bar is 10 \mum{}). Orange and blue lines delimit the edge of the InSe and \ws\ flakes.}
		\label{fig:01}
	\end{figure*}

\begin{figure*}
		\centering
		\includegraphics[width=\linewidth]{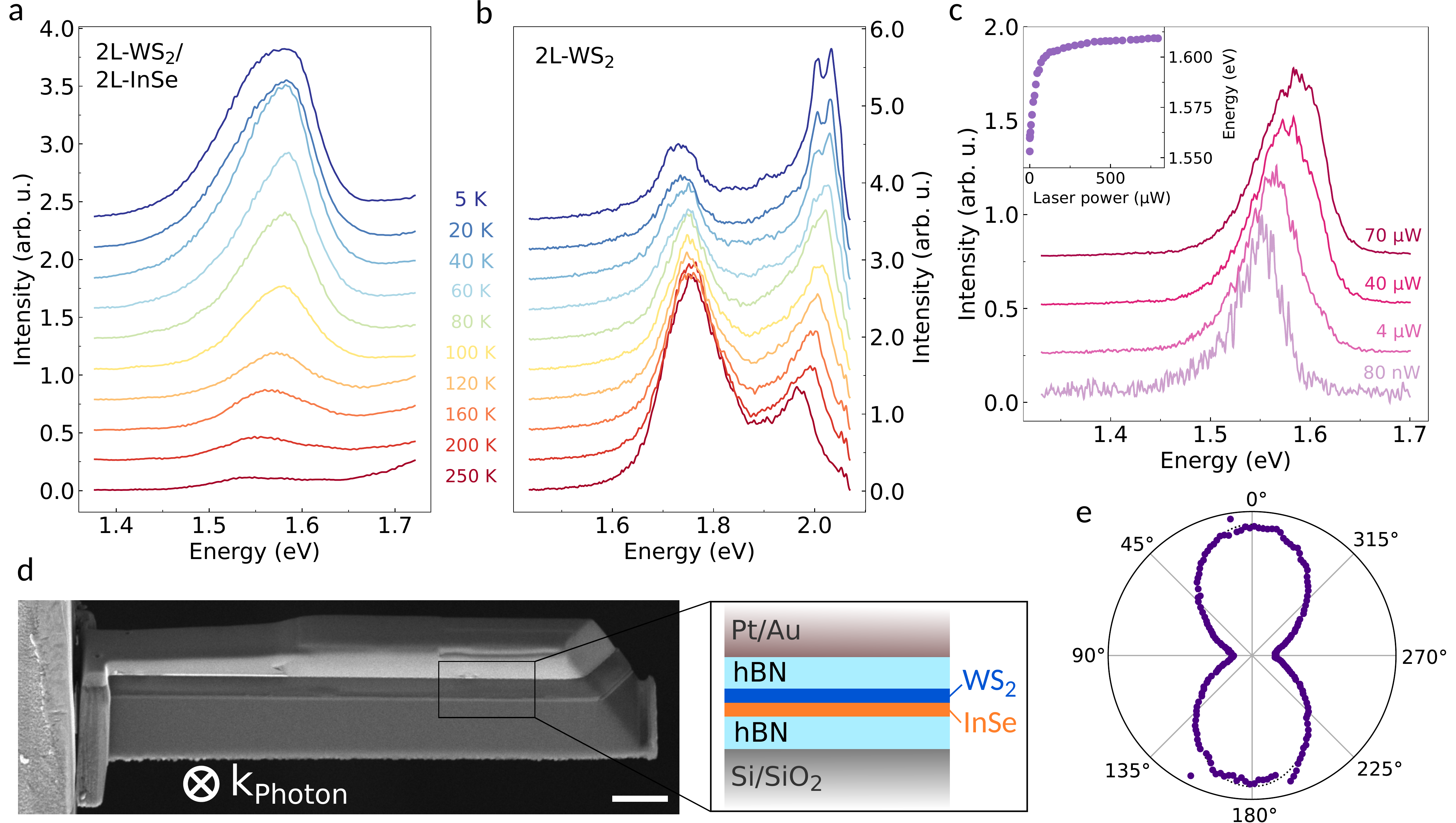}
		\caption{\textbf{Direct interlayer transition in 2L-InSe/2L-\ws{} interfaces.} \textbf{\sffamily a}. PL emission spectrum of a 2L-InSe/2L-\ws{} interface at different temperatures (curves offset for clarity). Upon lowering $T$ the peak intensity steadily increases as expected for a k-direct optical transition. For comparison, the PL emission spectrum of bare 2L-\ws\ is plotted in \textbf{\sffamily b} (curves offset for clarity). It shows that the 2.0 eV peak due to the k-direct transition at the K-point also increases upon cooling, whereas the intensity of the 1.73 eV peak originating from the k-indirect transition (from Q to \gam{}) decreases as $T$ is lowered. \textbf{\sffamily c}. PL spectra of a 2L-InSe/2L-\ws{} interface measured at $T =$ 5~K with different laser power (see indicated values; curves offset for clarity). The transition systematically blue-shifts upon increasing power (the full evolution of the peak position is shown in the inset, in the range between 80~nW - 800~$\mu$W). \textbf{\sffamily d}. Scanning electron microscope image of a lamella-shaped sample consisting of the layers shown in the right scheme (the interface consists of a 6L-InSe and a 2L-\ws{}). The lamella structure allows illumination with photons propagating in the plane of the interface, with an electric field polarized normal to the interface plane. The scale bar is 2\mum{}. \textbf{\sffamily e}. Polar plot of the PL of the interface in the lamella configuration, showing that the emitted light is predominantly polarized in the direction perpendicular to the interface (0\degree\ corresponds to a polarization perpendicular to the interface plane; the dashed line represents a fit of the data with a sinusoidal dependence).}
		\label{fig:02}
	\end{figure*}

	We perform photoluminescence (PL) measurements that allow the identification of spectral features in the light emitted by vdW interfaces originating from interlayer electron-hole recombination. The behavior representative of the systems that we have studied is illustrated in Fig.~\ref{fig:01} and \ref{fig:02}, with data measured on structures formed by bilayer InSe (2L-InSe) and bilayer \ws{} (2L-\ws{}), assembled to enable separate measurements of the PL coming from the individual layers and from their interface (see Fig.~\ref{fig:01}(a) for a schematic of the 2L-InSe/2L-\ws{} interface and Fig.~\ref{fig:01}(b) for the relevant aspects of the band structure). Lines of different color in Fig.~\ref{fig:01}(c) represent the PL measured at T = 5~K on 2L-\ws{} (blue line), 2L-InSe (orange line), and on their interface (purple line), and can be readily interpreted in terms of the known band structure of the materials forming the interface (see the arrows in Fig.~\ref{fig:01}(b)). The PL spectrum of 2L-\ws{} shows a (split) peak at approximately 2~eV originating from direct recombination of excitons and trions at the K-point, and a lower energy peak at 1.73~eV due to k-indirect recombination of electrons at the Q-point with holes at the \gam{}-point, as expected.\cite{zhao_evolution_2013,zhao_origin_2013} In the 2L-InSe spectrum a peak at approximately 1.9~eV is present, corresponding to the so-called A-transition in this system.\cite{mudd_tuning_2013,mudd_high_2015,mudd_direct--indirect_2016} The vdW interface PL, in contrast, is dominated by a peak close to 1.55~eV, significantly lower than the energy of the peaks identified in the spectra of the constituent materials, without any pronounced feature corresponding to those of 2L-\ws{} and 2L-InSe. We directly conclude that the interfacial PL cannot be explained in terms of intralayer transitions, and that the 1.55~eV peak originates from an interlayer transition resulting from charge transfer that quenches the PL of the individual layers.\\
	
	The temperature evolution of the interfacial PL intensity (Fig.~\ref{fig:02}\textbf{\sffamily a}) and its dependence on the excitation laser power (Fig.~\ref{fig:02}\textbf{\sffamily c}) do indeed exhibit the behavior characteristic of interlayer transitions that are direct in k-space. Fig.~\ref{fig:02}\textbf{\sffamily a} shows that upon reducing T from 250 to 5 K, the intensity of the 1.55~eV PL peak steadily increases, as expected for a transition that is direct in k-space. For comparison, Fig.~\ref{fig:02}\textbf{\sffamily b} shows that in 2L-\ws{} the amplitude of the 2~eV peak originating from k-direct recombination at the K-point also increases upon cooling, whereas the amplitude of the 1.75~eV peak due to the k-indirect transition between the Q and the \gam{} point decreases, as typical for phonon mediated processes. Upon increasing the excitation laser power, the transition energy increases by more than 50 meV before saturating as the power exceeds 100 $\mu$W (see Fig.~\ref{fig:02}\textbf{\sffamily c} and its inset). The blue-shift is a manifestation of the electrostatic potential generated by the “pumped” interlayer excitons, whose density increases (and eventually saturates) at higher excitation power. In simple terms, the photo-generated excitons consist of electrons residing in one layer (InSe) and of holes in the other layer (\ws{}), so that a higher exciton density results in a net electrostatic potential difference between the two layers and --owing to the interlayer nature of the transition-- in a shift of the transition energy (as discussed extensively in the literature --see for instance \cite{butov_magneto-optics_1999,laikhtman_exciton_2009}-- this interpretation in terms of an interlayer electrostatic potential difference is fully equivalent to accounting for the effect of the dipole-dipole interaction between the photo-excited excitons). In the same power range, a virtually identical behavior is observed in PL studies of interlayer excitons in vdW interfaces formed by \mose{} and \wse{} monolayers,\cite{rivera_observation_2015,nagler_interlayer_2017,miller_long-lived_2017} but it is never observed in individual monolayers. We conclude that the transition responsible for the 1.55~eV line in the PL power spectrum of the 2L-InSe/2L-\ws{} interface originates from an interlayer, k-direct transition, as expected for the \gam{}-\gam{} transition from the bottom of the conduction band of 2L-InSe to the top of the valence band of 2L-\ws{}.\\
	
	One more direct experimental indication of the origin of the interlayer transition that we observe comes from the analysis of polarization of the emitted light. At the \gam{}-point, the edge of the InSe conduction band and \ws{} valence band consist of atomic orbitals whose angular momentum component in direction perpendicular to the plane is zero.\cite{brotons-gisbert_out--plane_2019,hamer_indirect_2019} Theoretically this prescribes\cite{terry_infrared--violet_2018} that the polarization of the photons emitted in the interlayer transition should be perpendicular to the interface. To check if this expectation is satisfied we fabricated dedicated devices --by cutting into a lamella configuration  a block of 2~\mum{} x 20~\mum{} x 0.7~\mum{} out of h-BN encapsulated 2L-\ws{}/6L-InSe using  a focused ion beam (see Fig.~\ref{fig:02}(d))-- and measured the light emitted in the plane of the interface from the side of the lamella. The resulting polarization map is shown in Fig.~\ref{fig:02}(e), with a very pronounced out-of-plane photon polarization, as expected.\\
	
	On the basis of experimental observations similar to those just presented for 2L-InSe/2L-\ws{} interfaces, we conclude that also all other interfaces that we have investigated, based on combinations of thicker \ws\ and InSe multilayers, exhibit direct interlayer transition at \gam{}. The same is true for interfaces in which the \ws\ multilayers are substituted by \mose{}, or \mos{}. In all these systems, the PL spectrum of the interfaces exhibits a peak at an energy smaller than that of the spectral features of the individual constituent materials, whose amplitude increases upon cooling and whose frequency blue-shifts upon increasing the power of the exciting laser. The observed behavior is entirely consistent with the fact that all the semiconducting TMDs employed to assemble the interfaces have their valence band maximum centered at the \gam{}-point, and the same is true for the conduction band minimum of all the InSe multilayers.\cite{butov_magneto-optics_1999,laikhtman_exciton_2009,rivera_observation_2015,nagler_interlayer_2017} Interestingly, the PL of the interfaces can be even brighter than that of the individual constituents, showing that interlayer \gam{}-\gam\ transitions can result in an increased efficiency for light emission (see Supplementary Information section S3).\\
	
\begin{figure*}
		\centering
		\includegraphics[width=.64\linewidth]{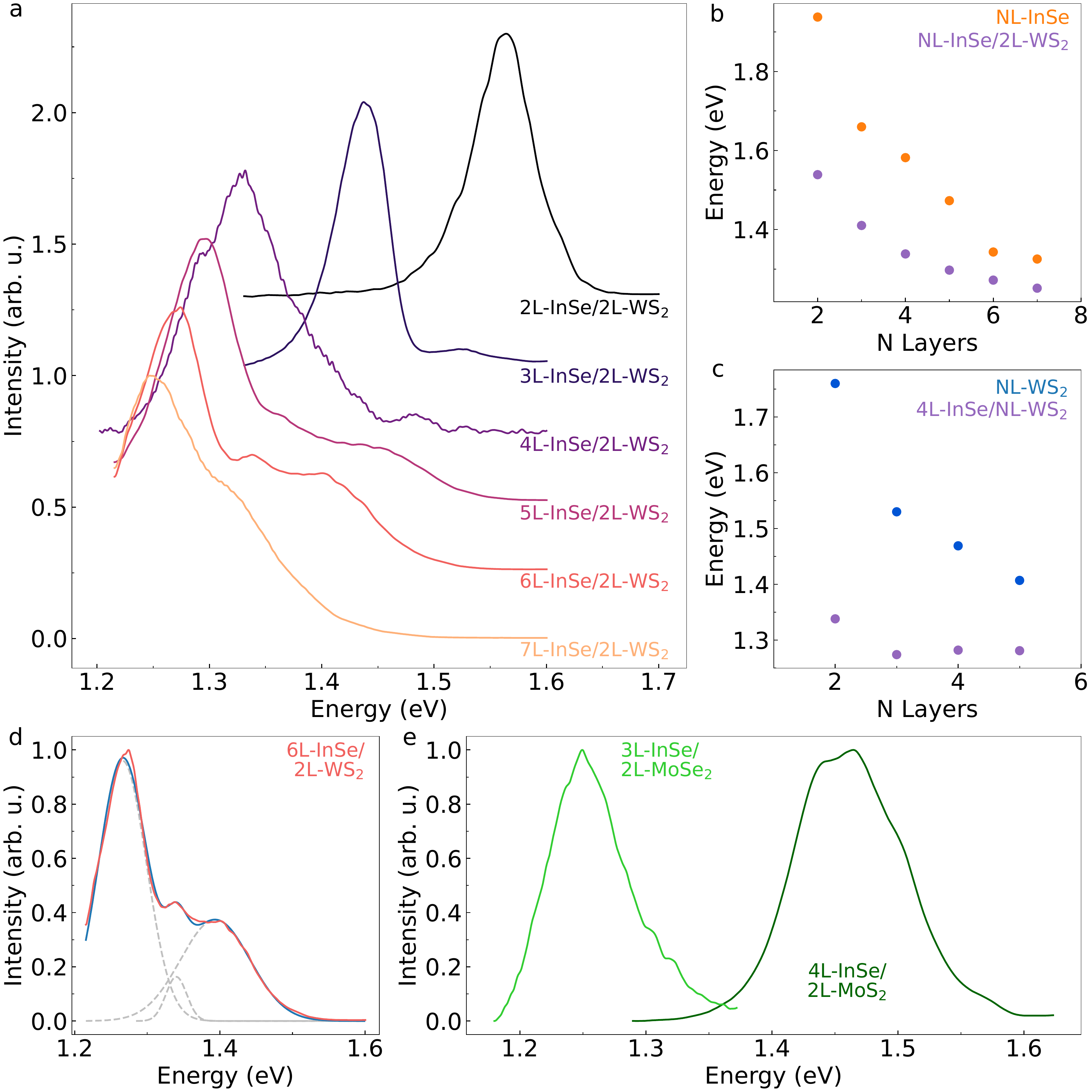}
		\caption{\textbf{Robust k-direct interlayer transitions at \gam{} in InSe-TMD multilayer interfaces.} \textbf{\sffamily a}. PL spectra of \emph{N}L-InSe/2L-\ws\, with $N$ varying between 2 and 7, measured at 5 K (curves are offset for clarity). In all cases a peak originating from a k-direct interlayer transition is observed. \textbf{\sffamily b}. Energy of the interlayer transition at \gam{} in \emph{N}L-InSe/2L-\ws{} interfaces (purple dots) as a function of $N$. The interlayer transition always occurs at energy lower than the transitions in the two constituent layers (the orange dots represent the intralayer transition energy in $N$L-InSe). \textbf{\sffamily c}. Interlayer transition energy in 4L-InSe/\emph{N}L-\ws{} as a function of the number $N$ of WS$_2$ layers (purple dots), extracted from the interfacial PL spectrum (the blue dots represent energy of the intralayer Q-\gam{} transition in \emph{N}L-\ws{}). \textbf{\sffamily d}. Decomposition of the PL spectrum of a 6L-InSe/2L-\ws\ interface: the measured data (red solid line) are reproduced (blue solid line) by summing multiple Gaussian lines (gray dashed lines) originating from the interlayer transition, the intralayer transition in 6L-InSe, and from an additional peak that we attribute to the hybridization of states at the valence band edge of 2L-\ws\ and  \emph{N}L-InSe (see main text). \textbf{\sffamily e}. PL spectra measured at  T = 5~K on interfaces realized with TMDs other than \ws\ (light green curve: 3L-InSe/2L-MoSe$_2$; dark green curve: 4L-InSe/2L-MoS$_2$). The observed peaks originate from k-direct interlayer transitions at \gam{}.}
		\label{fig:03}
	\end{figure*}

	Figure 3 shows representative data reproduced in more than 40 structures that we have investigated experimentally. In Fig.~\ref{fig:03}\textbf{\sffamily a} we show the evolution of the PL spectrum measured at T = 5~K on interfaces consisting of 2L-\ws{} and of InSe multilayers of increasing thickness (from 2L to 7L), and in Fig.~\ref{fig:03}\textbf{\sffamily b} we compare the thickness dependence of the interlayer transition energy extracted from Fig.~\ref{fig:03}\textbf{\sffamily a} (purple dots) to the energy of the intralayer transition in the corresponding InSe multilayers (orange dots). As stated above, for all thicknesses the interlayer transition occurs at a lower energy than the intralayer one. A similar behavior is observed upon fixing the thickness of the InSe layer and varying that of the \ws{} multilayers, as illustrated in Fig.~\ref{fig:03}\textbf{\sffamily c} for interfaces consisting of 4L-InSe and \emph{N}L-\ws{}, with \emph{N} varying from 2 to 5. Data measured on interfaces based on InSe and semiconducting TMDs other than \ws{} are presented in Fig.~\ref{fig:03}\textbf{\sffamily e}. Fig.~\ref{fig:03}\textbf{\sffamily e} shows the PL originating from interlayer electron-hole recombination in 3L-InSe/2L-\mose{}  and in 4L-InSe/2L-\mos{}: for these materials we did not systematically study the evolution of the interlayer transition energy upon varying thickness, but we did measure several interfaces combining multilayers of different thickness, and observed in all cases PL (again, with an amplitude that increases upon lowering temperature, at an energy that blue-shifts upon increasing excitation laser power). However, for interfaces based on \wse{} and InSe we did not observe any PL signal, despite the expected presence of a k-direct interlayer transition at the \gam{}-point. We believe that this is because the transition occurs at an energy of 0.8-0.9~eV, approaching the limit of sensitivity of our detector camera (the presence of a non-radiative recombination path that quenches the PL, \eg\ due to material-specific impurities creating in-gap states, can not be entirely ruled out at this stage).\\
	
    We conclude that  k-direct interlayer transitions at \gam{} are robust processes, as we have shown them to occur irrespective of the relative orientation of the multilayers forming the interface (in the majority of cases we did not align the crystals when assembling the structures), of a substantial lattice constant mismatch (approximatively 15~\%) between the constituents, and despite the fact that the band structure of TMD multilayers changes significantly upon varying their thickness. Besides substantiating our initial strategy to engineer systems for broad-spectrum optoelectronics, the ability to detect interlayer transitions in so many different interfaces enables the relative optical band alignment of entire classes of materials to be determined quantitatively, in a rather straightforward and reproducible way. This is a significant result, because band offsets of semiconductors are often complex to measure precisely, and different techniques give different results depending on details of how experiments are done.\\
	
\begin{figure*}
		\centering
		\includegraphics[width=\linewidth]{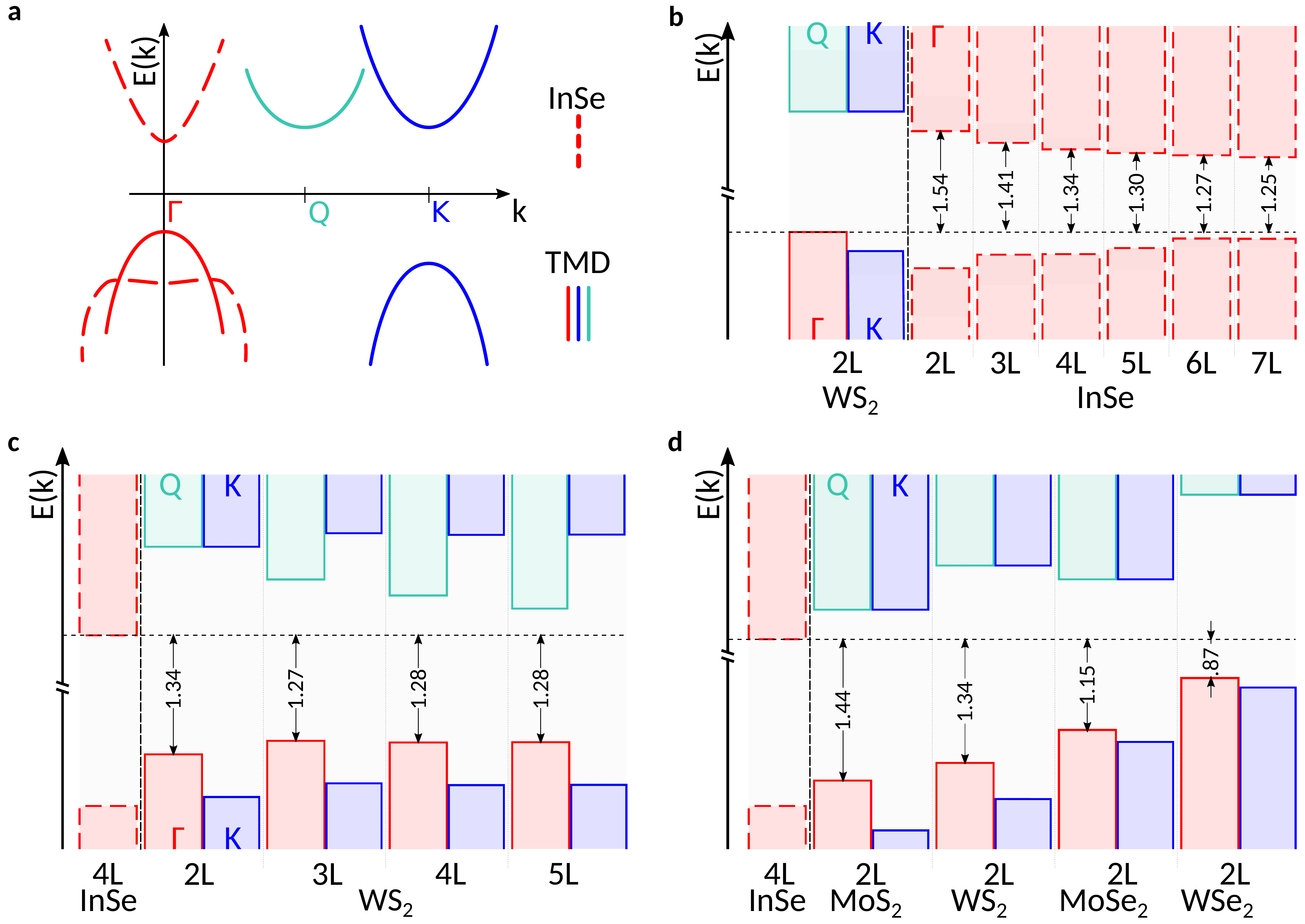}
		\caption{\textbf{Band diagaram of vdW interfaces.} \textbf{\sffamily a}. Relevant conduction and valence band edges of interfaces based on multilayers of InSe and semiconducting TMDs. Lines of the same color and style are used in the other panels of this figure to encode information about material and valleys: red lines indicates states at \gam{}, blue lines states at K, and green lines states at Q; if the line is continuous, the states belong to the TMD multilayers, if it is dashed to an InSe multilayer.  \textbf{\sffamily b}. Band alignment of \emph{N}L-InSe/2L-\ws{} interfaces, showing that the interfacial band gap of these systems is always formed by conduction band states in the InSe multilayers and valence band states in 2L-\ws{}, all centered around \gam{}. The size of the band  gap is indicated in the figure (in eV). \textbf{\sffamily c}. Same as \textbf{\sffamily b} for 4L-InSe/\emph{N}L-\ws{} interfaces. \textbf{\sffamily d}. Band alignment between 2L-TMDs and 4L-InSe obtained from optical data. For MoSe$_2$ we measured the PL on 3L-InSe/2L-MoSe$_2$ and reconstructed the alignment of 4L-InSe/2L-MoSe$_2$ using the data in \textbf{\sffamily b}, from which are known the relative alignment of the bands in 3L- and 4L-InSe.}
		\label{fig:04}
	\end{figure*}

	To understand how the alignment of the different band edges is determined, we discuss in detail the procedure for 2L-InSe/2L-\ws{} interfaces, whose relevant band edges are represented in Fig.~\ref{fig:04}\textbf{\sffamily a}. The interlayer transition occurring at 1.55~eV measures the distance in energy between the bottom of the 2L-InSe conduction band and the top of the 2L-\ws{} valence band at the \gam{}-point. The intralayer transition in 2L-InSe fixes the energy distance between the valence band maximum and the conduction band minimum (both near the \gam{}-point) in this 2D semiconductor (approximately 1.93~eV). Similarly, the indirect intralayer transition in 2L-\ws{} (approximately 1.73~eV) fixes the position of the bottom of the conduction band in bilayer \ws{} at the Q-point, relative to the maximum of the valence band at \gam{}. Since in 2L-\ws{} the conduction band edge at the Q and K points are nearly degenerate\cite{nguyen_visualizing_2019} (the energy difference is at most a few tens of meV, which we neglect here), we can use the k-direct, 2.0~eV intralayer transition at K in 2L-\ws{} to determine the energy of the maximum at K of the valence band of 2L-\ws{}. The relative alignment of all relevant band edges in 2L-InSe and 2L-\ws{} is then entirely determined. The same holds true for all other layers that we have investigated: 2L-\ws{} and InSe multilayers up to 7L-InSe (Fig.~\ref{fig:04}\textbf{\sffamily b}), for 4L-InSe and \ws{} of thickness increasing from 2L to 5L (Fig.~\ref{fig:04}\textbf{\sffamily c}), and for 4L-InSe combined with all the different semiconducting TMDs  (Fig.~\ref{fig:04}\textbf{\sffamily d}). We estimate the precision of the band-gap values extracted from this procedure to be 100~meV or better. A source of uncertainty comes from neglecting the binding energy of interlayer excitons, which is justified because for all interfaces investigated in our work this quantity is significantly smaller than 100 meV (see Section S4 of the Supplementary Information). What also limits the precision of our analysis are the assumption that the conduction band edges at the K and Q points of 2L TMDs are energy degenerate (correct only within a few tens of meV), and having neglected hybridization effects between the valence band edges of InSe and TMD multilayers in which these edges are energetically aligned (based on previous studies of other vdW interfaces reported in the literature,\cite{wilson_determination_2017} we estimate the phenomenon to cause an indetermination of 50 meV or less).\\
	
	The internal consistency of the band diagrams extracted from the measured interlayer and intralayer transition energies can be cross-checked directly with the measured data.  For instance, all measurements on interfaces formed by 2L-\ws{} and \emph{N}L-InSe shown in Fig.~\ref{fig:03}\textbf{\sffamily a} were performed with the excitation laser tuned at the K-K transition of 2L-\ws{}. The photo-excited electron at the K-point of the 2L-\ws{} conduction band has always enough energy to be transferred to the InSe multilayer (Fig.~\ref{fig:04}\textbf{\sffamily b}), which is why the interlayer transition in the PL spectrum is visible. For 5L-InSe and thicker layers, in addition, the top of the 2L-\ws{} K-valley becomes nearly degenerate with the top of the valence band of InSe at the \gam{}-point, so that also the hole in the K-valley of 2L-\ws{} can transfer into InSe. This has measurable consequences, since for 5L- and thicker InSe multilayers it leads to PL originating from the InSe intralayer transition, as well as to splitting of the transitions involving the \gam{}-point valence band edge (due to hybridization of states in 2L-\ws{} and in InSe). The PL spectra of 5L-, 6L-, and 7L-InSe (Fig.~\ref{fig:03}\textbf{\sffamily a}) do indeed show multiple peaks, one of which is due to the intralayer transition in InSe, and another that we attribute to the hybridization-induced splitting (see Fig.~\ref{fig:03}\textbf{\sffamily d} for the decomposition of the PL spectrum; a systematic discussion of hybridization effects will be presented elsewhere). Additional evidence in support of the band diagrams shown in Fig.~\ref{fig:04} is obtained from PL energy spectroscopy --i.e., from measurements of the PL intensity as a function of the exciting energy of the photon-- that we discuss in the supplementary information section S1. Finally, note also the different behavior of the band edges in InSe and \ws{} multilayers (compare Fig.~\ref{fig:04}\textbf{\sffamily b} and \ref{fig:04}\textbf{\sffamily c}): whereas in InSe both the conduction and valence band edges shift as thickness is increased (Fig.~\ref{fig:04}\textbf{\sffamily b}), in \ws{} the valence band edge remains nearly constant as the thickness is increased past that of bilayers (Fig.~\ref{fig:04}\textbf{\sffamily c}) in agreement with existing Angle-Resolved Photoemission Spectroscopy (ARPES) experiments.\cite{wilson_determination_2017}\\
	
	Besides demonstrating the validity of the proposed strategy for the realization of vdW interfaces supporting k-direct transitions, the band diagrams in Fig.~\ref{fig:04} show how the interfacial transition energy can be engineered by simply selecting different thicknesses of the most commonly available 2D materials. The heterostructures investigated here densely cover the interval between approximatively 1.0 and 1.6~eV, but  a much larger interval can be spanned by employing other known materials. Bilayers or multilayers of  MoTe$_2$, and of GaSe, for instance, will allow the energy interval to be extended further on the lower and higher end, respectively. It is this versatility and physical robustness that makes the use of interfacial transitions at the \gam{}-point ideally suited for the realization of broad spectrum optoelectronic applications. There is broad consensus that optoelectronics is one of the most promising domains for the development of devices based on interfaces of 2D materials, but the large-scale production and commercialization of such devices pose serious challenges as to the required level of material control. This is undoubtedly the case if device operation relies on a very fine control of the different constituents, such as a perfect alignment of 2D crystals, or combining materials with identical crystal lattices and lattice constants. While some of the challenges may be solved in the long term, these requirements are incompatible with virtually all large-area manufacturing techniques that are currently available. The results reported here, however, change the situation, since with an appropriate choice of materials vdW interfaces can be used as radiation sources covering a very broad frequency spectrum, in a mode of operation that is extremely robust against variations of interfacial structural details. This implies that even the simplest possible techniques to assemble large-area interfaces of atomically thin layers --such as liquid phase exfoliation followed by ink-jet printing-- can be employed for the scalable fabrication of structures with useful optoelectronic functionality.\\
	
	\section*{Methods}
	{\small%
		\noindent\textbf{Sample fabrication.} The fabrication of the heterostructures has been performed according to previous works (ref.~\cite{ponomarev_semiconducting_2018}) and reproduced here for completeness. The fabrication of the heterostructures used to perform the measurements discussed in the main text relies on conventional techniques that are commonly employed to manipulate atomically thin crystals, i.e., 2D materials. Atomically thin layers of TMDs  and InSe are obtained by mechanical exfoliation of bulk crystals in a nitrogen gas filled glove box with a < 0.5~ppm concentration of oxygen and water; the TMD crystals are purchased from HQ graphene and the InSe orignate from a Bridgman‐grown crystal of $\gamma$‐rhombohedral InSe. The exfoliated crystals are transferred onto Si/SiO$_2$ substrate and suitable flakes are identified by looking at their optical contrast under an optical microscope. The heterostructures are then assembled in the same glove box with a conventional pick-up and release technique based on either PPC/PDMS (Poly(propylene carbonate) /polydimethylsiloxane) or PC/PDMS (polycarbonate) polymer stacks placed on glass slides. The samples, encapsulated in between exfoliated h-BN crystals of few tens of nanometers, are removed from the glovebox and placed into the cryostat for optical investigations.\\
		\textbf{Optical measurements.}  Photoluminescence measurements in a backscattering geometry (illumination direction out-of-plane) are performed by using an optical microscope to illuminate the device with the incoming laser beam and to collect the resulting emitted light. The light source is a supercontinuum white light laser combined with a contrast filter, allowing to tune the laser wavelength between 400 and 1100~nm. The illumination wavelength for every spectrum is specified in the main text and the laser power kept at 50 $\mu{}W$, if not stated otherwise. All measurements are done with the sample placed in the vacuum chamber with a cryostat mounted on a piezoelectric driven x-y stage allowing positioning precision down to 50~nm (Cryovac KONTI). The laser beam is coupled onto the sample using an optical microscope with long working distance objectives producing a spot of approximately 1~\mum\ in diameter that can be focused anywhere on the sample surface. The light collected from the sample is sent to a Czerny-Turner monochromator with a grating of 150~gr/mm (Andor Shamrock 500i) and detected with a Silicon Charge Coupled Device (CCD) array (Andor Newton 970 EMCCD).\\
		\textbf{PL on lamella samples.} The fabrication of the lamella samples has been performed according to previous works (ref.~\cite{hamer_indirect_2019}) and reproduced here for completeness. Heterointerfaces for investigation of the in-plane PL are obtained in a similar manner as described in sample fabrication. However, an additional layer of thick (>100 nm) h-BN is transferred on top of the encapsulated heterointerface, followed by 3 nm AuPd + 5 nm amorphous carbon to protect the sample from ion damage. An area for extraction is then identified using AFM and covered with an additional protective platinum mask ($\approx$1 \mum\ thick) using a FEI Helios focused ion beam (FIB) dual-beam system. FIB milling is performed using a 30 kV Ga$^+$ beam to extract the selected area. An OmniProbe micromanipulator is used to lift the lamella, rotate it by 90\degree{}, and secure it onto an Omicron transmission electron microscopy grid. Finally, damaged edges of the lamella are polished by further FIB milling with a decreasing series of acceleration voltages (e.g. 5 kV, 47 pA and 2 kV, 24 pA). The final thickness of the specimen is <1 \mum{}, to suppress multiple internal reflections of the emitted light. The polarization measurements are taken at 4~K in an AttoDry 100 cryostat (AttoCFM inset), using Princeton Instruments Acton Spectrapro SP-2500i CCD with 300 gr/mm grating.
	}
	
	\section*{Acknowledgements}
	We gratefully acknowledge Alexandre Ferreira for continuous and precious technical support. AFM gratefully acknowledges financial support from the Swiss National Science Foundation (Division II) and from the EU Graphene Flagship project. NU gratefully acknowledges financial support from the Swiss National Science Foundation through the Ambizione program. RVG and VIF gratefully acknowledges financial support from European Graphene Flagship Core 2 Project under grant agreement 785219, ERC Synergy Grant Hetero2D, EPSRC grants EP/S030719/1, EP/S019367/1, EP/P026850/1 and EP/N010345/1, Lloyd Register Foundation Nanotechnology Grant. K.W. and T.T. acknowledge support from the Elemental Strategy Initiative conducted by the MEXT, Japan, A3 Foresight by JSPS and the CREST (JP-MJCR15F3), JST. ZDK acknowledges the support from the National Academy of Sciences of Ukraine
	
	\section*{Author contributions}
	{\small N.U., E.P., J.Z., and D.T. have equally contributed. AFM and VIF have conceived the idea of this project. N.U., E.P., D.T., D.D., J.Z., J.H., and I.G.L. fabricated heterostructures, participated in their characterization, and analyzed the data. D.T., J.Z., J.H., A.Z, and R.V.G. designed and performed the experiment on the lamella samples. V.Z. and V.I.F. developed the theory of optical transitions at \gam{} in vdW  interfaces. Z.R.K., Z.D.K., A.P., T.T., and K.W. provided InSe and h-BN crystals. N.U., E.P., R.V.G., V.I.F, and A.F.M wrote the manuscript with input from all authors. All authors discussed the results.}
	
	\section*{Competing financial interests}
	The authors declare no competing financial interests.
	
	\section*{Data availability}
	The data that support the findings of this study are available from the corresponding authors without any restriction.
	

\end{document}